\documentclass[%
12pt,T
oneocolumn,
nofootinbib,
amsmath,amssymb,
aps,
floatfix,
unsortedaddress
]{revtex4-2}

\usepackage{graphicx,color}
\usepackage{subfig}
\usepackage{dcolumn}
\usepackage{bm}
\usepackage{graphicx,color}
\usepackage{float}
\usepackage{multirow}

\usepackage{amsmath}
\usepackage{mathtools}

\usepackage{cancel}
\usepackage[pdfpagelabels, pdfencoding=auto, psdextra]{hyperref}
\hypersetup{%
	pdfsubject=Paper,
	pdfkeywords={nuclear physics} {Pionless EFT} {Two-Nucleon} {Lattice} {L\"usher formula},
	unicode = true,
	breaklinks = true,
	colorlinks = true,
	linkcolor = blue,
	citecolor = blue,
	menucolor = blue,
	citecolor = blue,
	urlcolor = blue
}
\graphicspath{{./}}
\begin{document}
	
	\title{Probing $NN\Omega_{ccc}$ three-body systems with the modern QCD $N\Omega_{ccc}$ interaction }
	\author{Faisal Etminan}
	\email{fetminan@birjand.ac.ir}
	\affiliation{ Department of Physics, Faculty of Sciences, University of Birjand, Birjand 97175-615, Iran
	}%
	\affiliation{ Interdisciplinary Theoretical and Mathematical Sciences Program (iTHEMS), RIKEN, Wako 351-0198, Japan}
	\author{Lucas Happ}
	\email{lucas.happ@riken.jp}
	\affiliation{ Few-body Systems in Physics Laboratory, RIKEN Nishina Center, Wakō, Saitama 351-0198, Japan }%
	

	\date{\today}%
	\begin{abstract}
		Newly, first-principles lattice QCD results at the physical pion mass,  $ m_\pi \backsimeq 137.1 $ MeV, have been reported by the HAL QCD Collaboration for the S-wave interaction between the nucleon ($N$) and the triply charmed Omega baryon ($\Omega_{ccc}$). The $N\Omega_{ccc}$ potentials in the spin-1 $ \left(^{3}S_{1}\right) $ and spin-2 $ \left(^{5}S_{2}\right) $ channels were derived and found to be attractive, though no two-body bound state was supported in these channels. 
		The present work investigates the $NN\Omega_{ccc}$ three-body system
		using the Malfliet-Tjon $NN$ potential. Analyses of spin-1, spin-averaged, and spin-2 $N\Omega_{ccc}$ channels (at Euclidean times 16, 17, 18) reveal a three-body bound state only for the d-$\Omega_{ccc}$ configuration with spin $(0)1/2^{+}$ and $t/a=16$. Its binding energy ($B_3 = -2.255$ MeV) lies slightly below the deuteron's ($B_d = -2.23$ MeV). 
		Other parameter sets do not yield a bound state, 
		and complex scaling analysis indicates these configurations correspond to virtual states rather than resonances.
		The Coulomb potential's role was also examined to differentiate charged states. 
	\end{abstract}
	\maketitle

	\section{Introduction} \label{sec:intro}	
	The accurate determination of binding energies for heavier hypernuclei is expected to enhance understanding of hyperon–nucleon ($YN$) interactions in many-body environments. In nature, several well-understood few-body systems involving nucleons are expected to reinforce binding through their interactions, with a general trend that binding energy per nucleon ($ B/A $) becomes stronger as the nucleon number $ A $ increases in ordinary (non-strange) nuclei. A similar trend is observed in the strange sector. Although definitive evidence for -1 strangeness dibaryon states remains elusive, bound hypertriton states, such as $_{\Lambda}^{3}\textrm{H}\ (I=0, J^{P}=1/2^{+})$, $_{\Lambda}^{4}\textrm{H}\ (I=0, J^P = 0^{+})$, and $_{\Lambda}^{5}\textrm{He}\ (I=0, J^{P}=1/2^{+}) $, have been experimentally confirmed, with separation energies of approximately $130 \pm 50$ keV, $2.04 \pm 0.04$ MeV and $3.12\pm0.02$ MeV, respectively. Moreover, the -3 strangeness dibaryon state in the $N\Omega$ $ ^{5}S_{2} $ channel is supported by a central binding energy of 1.54 MeV~\cite{etminan2014,Iritani2019prb}, while the d-$ \Omega $ system in the maximal spin state $(I)J^{\pi} = (0)5/2^{+}$ is bound with a binding energy of about 20 MeV~\cite{Garcilazo2019}.

	Beyond two-body systems, the role of three-body forces is further illustrated by recent discussions prompted by measurements of the proton–deuteron correlation function by the ALICE collaboration~\cite{PhysRevX.14.031051}, wherein two-body interactions alone could not adequately fit the data, prompting debate on the necessity of including three-body or higher-order interactions~\cite{torresrincon2025femto-pd}. These examples, together with a substantial body of studies, demonstrate the importance of the three-body problem~\cite{RevModPhys.41.497,Hiyama2020, Filikhin2024prd,etminanPRC2025-112-4, etminanPRC2025-112-6}.
	
	As a natural extension, the investigation of charmed hypernuclei is of interest, following the early work~\cite{PhysRevLett.39.1506} conducted after the discovery of the $\Lambda_{c}$ baryon. With charm quarks included, a one-boson-exchange potential model for $NY_{c} $ $ \left(\textrm{Y}_{c}=\Lambda_{c},\Sigma_{c}\right) $ was developed~\cite{PhysRevD.16.799}, and the possibility of both $ \Lambda_{c} $ and $ \Sigma_{c} $ nuclear bound states was predicted for heavy nuclei~\cite{PhysRevLett.39.1506, Miyamoto2018}.
	Moreover, the chromopolarizabilities of fully-heavy baryons $\Omega_{QQQ^{\prime}}\left(Q,Q^{\prime}=b,c\right)$, which determine the interaction strength between a fully-heavy baryon and a nucleon is studied in the framework of potential nonrelativistic quantum chromodynamics~\cite{PhysRevD.107.034020, WU2025}.
	
	Among various two-hadron systems, di-hadrons with distinct quark-flavor content have been explored using the time-dependent HAL QCD method~\cite{ishii2012}, including $N \Omega_{sss} $~\cite{Iritani2019prb}, $N \phi $~\cite{yan2022prd}, $N J/\psi (N\eta_c) $~\cite{LyuPLB2025}, $N\Omega_{ccc}$~\cite{Zhang-NOmegaccc}. These systems are of particular interest because they offer a unique opportunity to investigate long-range and short-range interactions in the absence of quark Pauli blocking.
	
	Recently, a first lattice QCD simulation of the $N\Omega_{ccc}$ potentials in the $ ^{3}S_{1} $ and $ ^{5}S_{2} $ channels at the physical pion mass has been published by the HAL QCD Collaboration~\cite{Zhang-NOmegaccc}. In this work, the spin-1 and spin-2 $N\Omega_{ccc}$ potentials are employed in conjunction with the realistic $NN$ MT potential in the $ ^{3}S_{1} $ channel to study d-$\Omega_{ccc}$ in the $(I)J^{\pi} = (0)1/2^{+}$ and $(0)5/2^{+}$ channels.  It is noted that these channels cannot couple to other channels, as discussed in Ref.~\cite{Zhang-NOmegaccc}, in contrast to the $N \Omega_{sss} $ system, which lies above the $ \Lambda\Xi $ and $ \Sigma\Xi $ thresholds. The $N\Omega_{ccc}$ threshold, approximately 5740 MeV, is well below both the $ \Lambda_{c}\Xi_{cc} $ (5910 MeV) and $ \Sigma_{c}\Xi_{cc} $ (6080 MeV) thresholds, making it the lowest state among all charmed dibaryon systems with charm number three. Such an energy, about 200 MeV below these thresholds, provides an ideal, uncontaminated setting to examine the low-energy $N\Omega_{ccc}$ interactions.
	
	Filikhin et al.~\cite{filikhin2025omega3cnn,FilikhinPRC2025OmegaNuclei} recently used the Faddeev formalism to analyze the $NN\Omega_{ccc}$ system, employing HAL QCD S-wave $N\Omega_{ccc}$ potentials while neglecting Coulomb forces. Their study revealed no bound states for the $NN\Omega_{ccc}$ system. Instead, near-threshold resonances were predicted for the $J^{\pi}=5/2^{+}$ and $1/2^{+}$ states, with resonance energies located 1.1 MeV below and 0.0 MeV at the three-body breakup threshold, respectively, as observed at Euclidean time $t/a = 16$.

	Certain resonances cannot be fully explained by the traditional quark model, hinting at a more complex underlying structure. These resonances often involve hadrons with substructures that differ from simple quark–antiquark mesons or three-quark baryons. Such findings challenge the original quark model, which otherwise effectively describes most low-mass hadrons~\cite{FilikhinPRD2020resonance}.  
	
	\section{Two-body potentials} \label{sec:Two-body-potentials}
	We employ the lattice HAL QCD $N\Omega_{ccc}$ potential in the spin-1 and spin-2 channels, as published very recently~\cite{Zhang-NOmegaccc}. The $N\Omega_{ccc}$ potential in the S wave is extracted from lattice data at  $ t/a = 16, 17 $  and $ 18 $, with $ a\simeq0.084372 $ fm or equivalently $ a^{-1}\simeq2338.8 $ MeV, yielding a spatial extent $ L\simeq8.1 $ fm sufficient to accommodate two-baryon interactions. To calculate physical observables, uncorrelated fits were performed on the lattice QCD extracted potential using a phenomenological two-range Gaussian form
	\begin{equation}
		V^{fit}\left(r\right)=\sum_{i=1}^{2}\alpha_{i}e^{-\left(r/\beta_{i}\right)^{2}}. \label{eq:pot_nOmegaccc}
	\end{equation}
	The potential in the range $ 0 < r < 3 $ fm was fitted. The fitting parameters used in the calculations are given in Table 2 of  Ref.~\cite{Zhang-NOmegaccc}.  No two-body bound state is observed for these interactions.
	
	In addition to calculations for the deuteron-$\Omega_{ccc}$ interaction in the channels \((I)J^{\pi}=(0)1/2^{+}\) (minimum spin) and \((0)5/2^{+}\) (maximum spin), the interaction in an intermediate spin configuration channel \((I)J^{\pi}=(0)3/2^{+}\) can be defined as
	$ V_{N\Omega_{ccc}}(^{4}S_{3/2})=\frac{5}{8}\,V_{N\Omega_{ccc}}(^{5}S_{2})+\frac{3}{8}\,V_{N\Omega_{ccc}}(^{3}S_{1})$.
	Such a definition is intended to offer insights into the underlying spin-coupling mechanisms and to reveal patterns or symmetries in the interaction strengths, potentially leading to broader conclusions about the \({N\Omega_{ccc}}\) system. Exploration of this channel is expected to bridge theoretical predictions with experimental possibilities.

	For $NN$ interactions, the  Yukawa-type Malfliet-Tjon potential is employed~\cite{MALFLIET1969161, FriarPRC1990},		
	\begin{equation}
		V_{NN}\left(r\right)=\sum_{i=1}^{2}C_{i}\frac{e^{-\mu_{i}r}}{r}, \label{eq:VNN}
	\end{equation}
	The parameters  $ C_{i} $ and $ \mu_{i} $ are given in Table 1 of Ref.~\cite{etminan2024prc}. This potential reproduces the deuteron binding energy of $-2.2307$ MeV. In the present work, the results have been calculated with  
	$m_{N} = 938.9$ $\textrm{MeV}/\textrm{c}^{2}$ and $m_{\Omega_{ccc}} = 4796.8$ $\textrm{MeV}/\textrm{c}^{2}$.

		To differentiate among the charged states of the $NN\Omega_{ccc}$ system, the calculations are carried out for the Coulomb interaction as:
	\begin{equation}
	V_{\text{Coul}}(r)=\alpha_{f}Z_{1}Z_{2}\dfrac{e^{-r/r_{0}}}{r}. \label{eq:Coul}
	\end{equation}
	with $Z_{1}Z_{2}$ being the product of the nuclear charges of particle $1$ and $2$ ( $Z_{p^{+}}Z_{p^{+}}=1$ and $Z_{p^{+}}Z_{\Omega_{ccc}^{++}}=2$).
	Here, \(\alpha_f = 1/137.036\) is the fine-structure constant, and \(r_0 = 50\) fm is a screening radius \cite{Garcilazo2019}.

	\section{Method}
	We describe three-body systems in the center-of-mass frame via Jacobi coordinates. In this scheme, the full system is characterized by two relative coordinates: $\vec{r}$, representing the distances between each pair of particles, and $\vec{R}$, denoting the distance between the center of mass of the pair and the corresponding third particle. An important aspect is the existence of three equivalent sets of Jacobi coordinates, each corresponding to a distinct partition of the three particles into a pair and a single particle. Consequently, the notation is commonly defined as follows:	
	\begin{equation}
		\begin{array}{cc}
			\vec{r}_{i}\equiv\vec{s}_{j}-\vec{s}_{k}, & \vec{R}_{i}\equiv\vec{s}_{i}-\frac{m_{j}\vec{s}_{j}+m_{k}\vec{s}_{k}}{m_{j}+m_{k}},\end{array}
	\end{equation}
	where $m_i$ and $\vec{s}_i$ respectively denote the mass and absolute coordinate of particle $i$, and the indices $i,j,k$ cyclically traverse $\left(1,2,3\right)$.
	The Gaussian Expansion Method (GEM) for three-body systems is outlined below, focusing on the case of central forces alone. We consider the Schr\"odinger equation	
	\begin{equation}
		\left[-\frac{\hbar^{2}}{2\mu_{ij}}\nabla_{\vec{r}_{k}}^{2}-\frac{\hbar^{2}}{2\mu_{k}}\nabla_{\vec{R}_{k}}^{2}+V_{12}\left(\vec{r}_{12}\right)+V_{23}\left(\vec{r}_{23}\right)+V_{31}\left(\vec{r}_{31}\right)\right]\Psi\left(\vec{r}_{k},\vec{R}_{k}\right)=E\Psi\left(\vec{r}_{k},\vec{R}_{k}\right),
	\end{equation}	
	where the reduced masses are defined by $\mu_{ij}=\frac{m_{i}m_{j}}{m_{i}+m_{j}}$ and $\mu_{k}=\frac{\left(m_{i}+m_{j}\right)m_{k}}{m_{i}+m_{j}+m_{k}}$. The total three-body wave function is expressed as $\Psi\left(\vec{r},\vec{R}\right)=\sum_{i=1}^{3}\Psi^{\left(i\right)}\left(\vec{r}_{i},\vec{R}_{i}\right)$, where the components $\Psi^{\left(i\right)}$ are functions of three distinct sets of Jacobi coordinates. Each Faddeev component is expanded in terms of Gaussian basis functions:	
	\begin{equation}
		\Psi^{\left(i\right)}\left(\vec{r}_{i},\vec{R}_{i}\right)=\sum_{\zeta}A_{\zeta}\phi_{\zeta}\left(\vec{r}_{i}\right)\Phi_{\zeta}\left(\vec{R}_{i}\right),
	\end{equation}	
	with $\phi_{\zeta}\left(\vec{r}\right)=N_{l,m}r^{l}e^{-\nu_{n}r^{2}}Y_{l,m}\left(\theta,\phi\right)$ and $\Phi_{\zeta}\left(\vec{R}\right)=N_{L,M}R^{L}e^{-\lambda_{N}R^{2}}Y_{L,M}\left(\theta,\phi\right)$. Here, $\zeta\equiv\left\{ n,l,N,L\right\} $ represents the quantum numbers of each component, $N_{l,m}$ is the normalization factor, and $Y_{l,m}$ denotes the spherical harmonics. The Gaussian ranges $\nu_{n}=1/r_{n}^{2}$ and $\lambda_{N}=1/R_{N}^{2}$ are   
	defined in a geometric progression, where $r_{n} = r_{1}a^{n-1}$ (and $R_{n} = R_{1}A^{N-1}$), with $n$ ranging from 1 to $n_{\text{max}}$ (and $N$ ranging from 1 to $N_{\text{max}}$)~\cite{HIYAMA2003223}. 	
	The parameters $a$ and $A$ are defined by choosing the minimum and maximum ranges. The three-body Schr\"odinger equation was solved using the FewBodyToolkit.jl software package~\cite{happ2025fewbodytoolkit}.

	Near-threshold bound or resonant states of the three-body system can be determined using the complex scaling method (CSM)~\cite{kukulin2013theory,AoyamaPTP2006,happ2025stab}. The CSM provides a direct approach for calculating the energies and decay widths of resonant states by performing an analytical continuation of the Schr\"odinger equation. This is achieved by applying a complex rotation $U\left(\theta\right)$ to the radial coordinate $r$ and momentum $p$, such that $U\left(\theta\right)r=re^{{\rm i}\theta}$ and $U\left(\theta\right)p=pe^{-{\rm i}\theta}$, respectively. The Hamiltonian is transformed as $H\left(\theta\right)=U\left(\theta\right)HU\left(\theta\right)^{-1}$, and the corresponding Schr\"odinger equation, $H\Psi=E\Psi$, is transformed into:	
	\begin{equation}
		H\left(\theta\right)\Psi\left(\theta\right)=W\Psi\left(\theta\right),
	\end{equation}	
	where $\Psi\left(\theta\right)=U\left(\theta\right)\Psi=\exp\left(3{\rm i}\theta/2\right)\Psi\left(re^{{\rm i}\theta}\right)$, with a complex eigenvalue $W\left(\left|arg\:W\right|\leq2\theta\right)$.
	
	As a result, the complex-scaled Schr\"odinger equation yields the binding energy $E_{B}$ as the real eigenvalue, which remains unchanged under complex scaling. For resonant states, the resonance energy $E_{r}$ and the decay width $\Gamma$ are obtained as the complex eigenvalues $W=E_{r}-{\rm i}\Gamma/2$ for sufficiently large rotation angles $\theta$.	
	\section{Numerical Results} 	\label{result}	
	The $N\Omega_{ccc}$ two-body system was analyzed for all channels (spin-1 $\left(^{3}S_{1}\right)$, spin-averaged, and spin-2 $\left(^{5}S_{2}\right)$) at  Eucludian times $\left(16,17,18\right)$, using the parametrizations provided in Table 2 of Ref.~\cite{Zhang-NOmegaccc}. Although the interaction was found to be attractive, it supports no bound state or resonance in the two-body system.

	We perform a variation of the coupling constant~\cite{kukulin2013theory} via $	V_{N\Omega_{ccc}}^{\gamma}(r) = (1+\gamma)\,V_{N\Omega_{ccc}}(r)$.
	For $\gamma>0$, the system is artificially more strongly bound, and for $\gamma=0$ the physical situation is recovered. This variation assists in estimating the results at the physical point based on stable bound-state calculations, which are predominantly unattainable there. To provide a conservative estimate of the real part of the three-body energies at the physical point, a linear extrapolation down to $\gamma=0$ is performed. The actual values are expected to lie below the extrapolated estimates. We highlight that when approaching thresholds through this coupling-constant variation, states may first become virtual states before turning into resonances.

	The CSM can reveal resonances, provided that the rotation angle is sufficiently large to separate the resonances from the corresponding continua. 
	The theoretical maximum rotation angle for this method is \(45^{\circ}\). In the following, calculations were performed for a large rotation angle of \(\theta =30^{\circ}\). 
	Increasing this angle further did not reveal new results, however numerical instabilities appeared near $\theta = 40^{\circ}$.

	\subsection{$nn\Omega_{ccc}$ system}
\begin{figure*}[hbt!]
		\centering
		\includegraphics[scale=0.4]{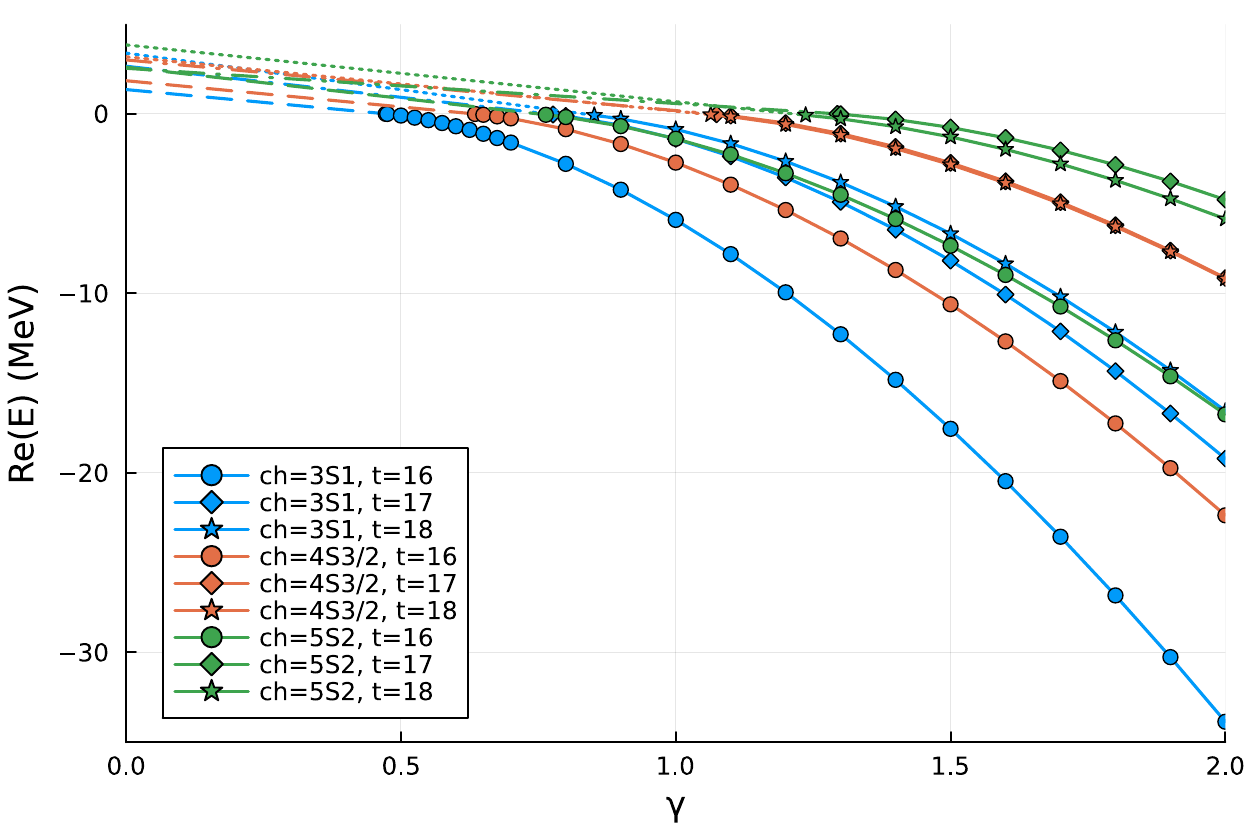}
		\caption{Coupling constant variation for the $nn\Omega_{ccc}$ system.\label{fig:nnOmega_ccv}}
	\end{figure*}

\begin{figure*}[hbt!]
		\centering
		\includegraphics[scale=0.4]{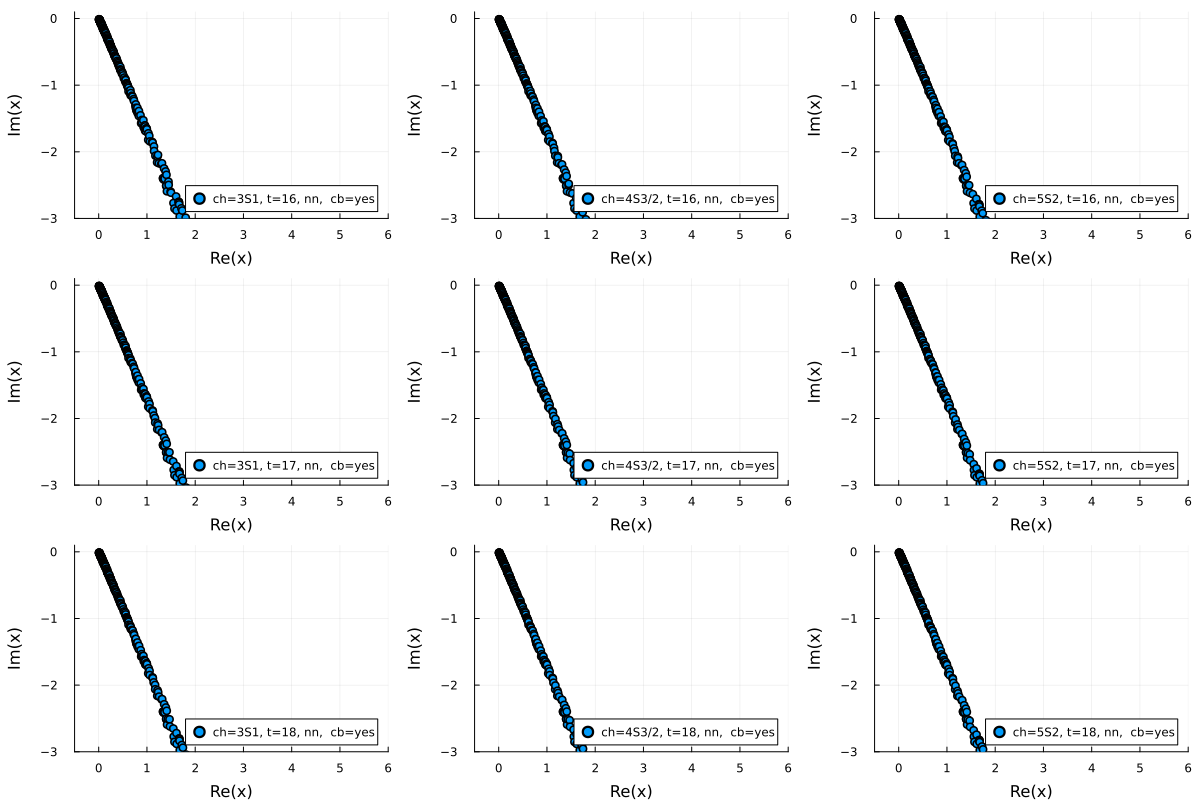}
		\caption{Complex-scaled spectrum for the $nn\Omega_{ccc}$  system. No resonance is found. 
			Since this system contains no bound two-body subsystem, there is only a single rotated continuum starting at the origin.
			\label{fig:nnOmega_csm}}
	\end{figure*}

	In this system, the Coulomb interaction is absent and therefore plays no role.  The variation of the coupling constant for the $nn\Omega_{ccc}$ system is illustrated in Fig.~\ref{fig:nnOmega_ccv}.
	The extrapolated three-body energies of the $nn\Omega_{ccc}$ state (relative to the three-body breakup threshold) at $\gamma=0$ are presented in Table~\ref{tab:nnOmega}. These values are obtained by employing the $N\Omega_{ccc}$ interaction across three channels: $^{3}S_{1}$, ${}^{4}S_{3/2}$, and ${}^{5}S_{2}$, at Euclidean time slices $t/a = 16, 17, 18$.
	Furthermore, the corresponding complex-scaled spectrum is illustrated in Fig.~\ref{fig:nnOmega_csm}.

	The variation of the coupling constant yields no bound state in the $nn\Omega_{ccc}$ system, but possible resonances in the range $1-4$ MeV. However, the complex scaling does not reveal any resonances in this range. Therefore, the system either possesses very broad resonances, or virtual states, which  cannot be directly identified by the CSM.  
	Moreover, we should note that for some configurations the extrapolation to $\gamma=0$ starts far away from that value, which may limit the precision of these estimates.
	
	\begin{table}
		\caption{Three-body energies (in MeV) of the $nn\Omega_{ccc}$ system with respect to the three-body breakup threshold.}
		\centering
		\begin{tabular}{c|ccc c ccc c ccc }
			\hline
			Channel	& \multicolumn{3}{c}{$ 16 $}&& \multicolumn{3}{c}{$17$}&&  \multicolumn{3}{c}{$18$}\\
			\cline{1-1} \cline{2-4} \cline{6-8} \cline{10-12}
			$^{3}S_{1}$   & 1.30 &  &  && 2.63 &  &  && 3.37 &  &  \\  
			$^{4}S_{3/2}$ & 1.83 &  &  && 2.98 &  &  && 3.16 &  &  \\  
			$^{5}S_{2}$   & 2.59 &  &  && 2.47 &  &  && 3.83 &  &  \\  			
		\end{tabular}
		\label{tab:nnOmega}
	\end{table}	
	
	\subsection{$pp\Omega_{ccc}$ system}
\begin{figure*}[hbt!]
		\centering
		\includegraphics[scale=0.35]{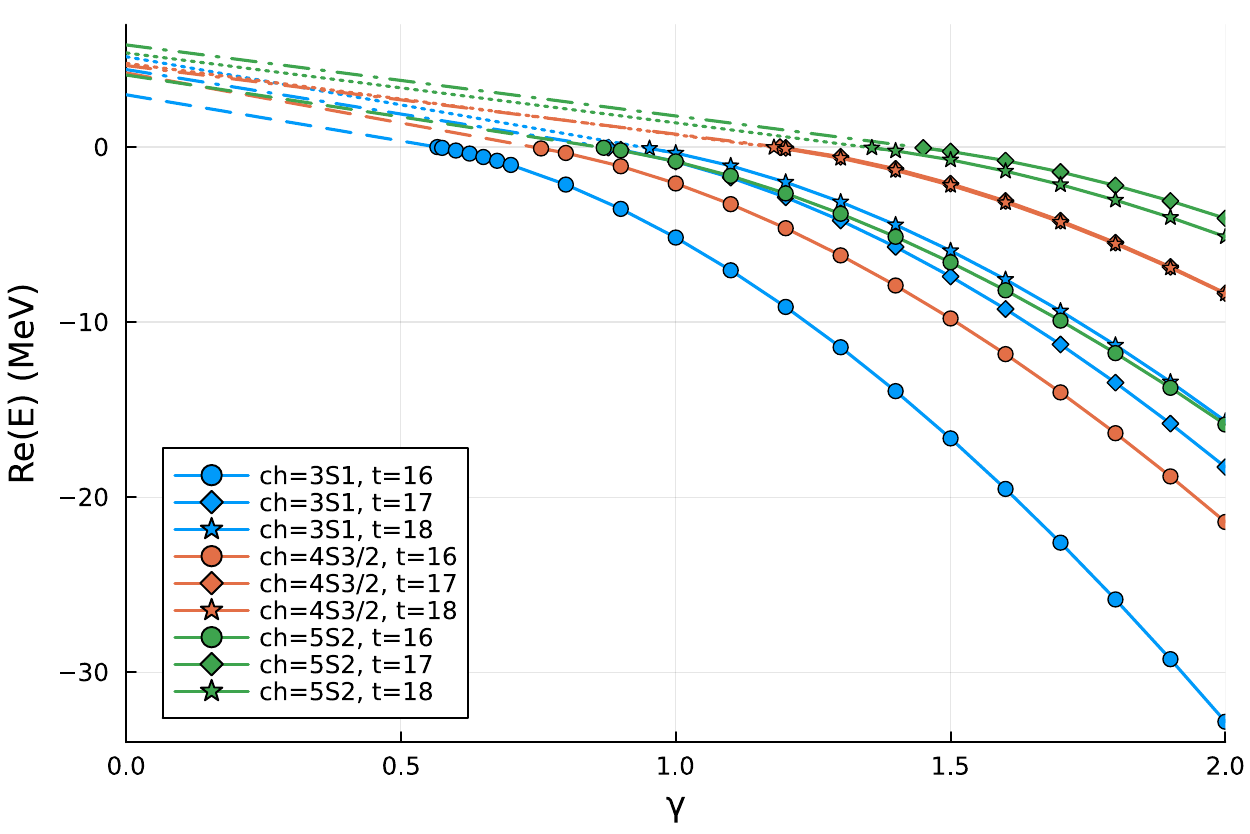} 	
			\includegraphics[scale=0.35]{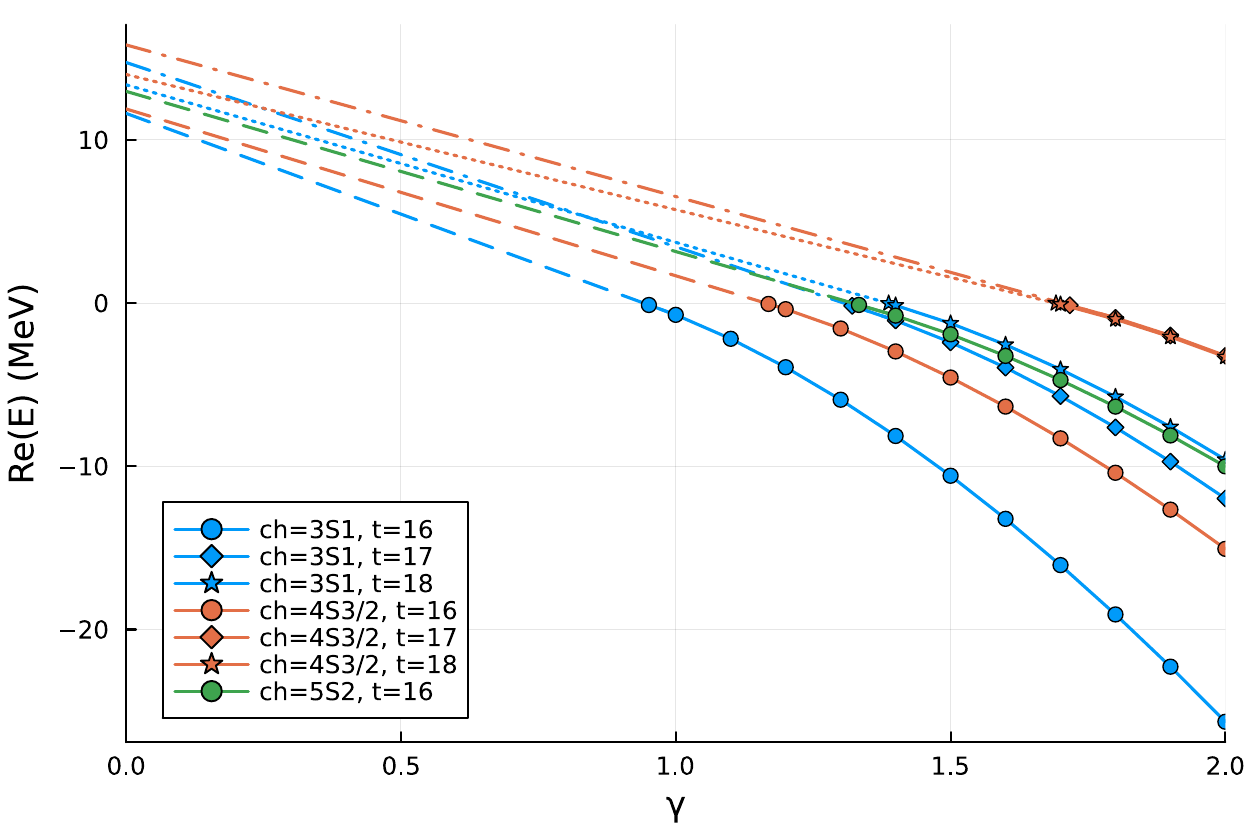}
		\caption{Coupling constant variation for the $pp\Omega_{ccc}$ system, without (left), and with (right) Coulomb interaction.\label{fig:ppOmega}}
	\end{figure*}

\begin{figure*}[hbt!]
		\centering
		\includegraphics[scale=0.4]{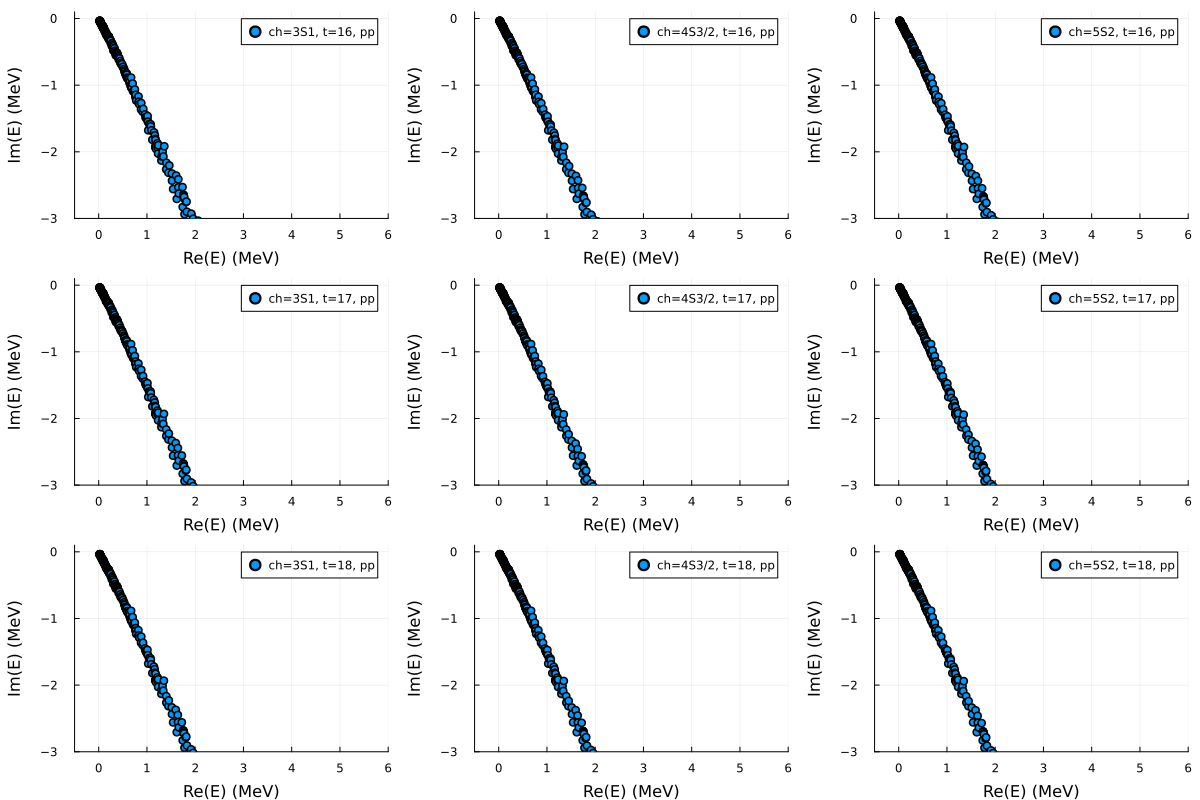}
		\caption{Complex-scaled spectrum for the $pp \Omega_{ccc}$  system. No resonance is found. 
			 As for the $nn\Omega_{ccc}$ case, this system has no bound two-body subsystem, hence there is only a single rotated continuum starting at the origin.
			\label{fig:ppOmega_csm}}
	\end{figure*}

	 Due to the excessive repulsion in this system, the resulting three-body energies are significantly larger than for the $nn\Omega_{ccc}$ system. 
	Accordingly, the extrapolation down to $\gamma=0$ starts further away and is therefore less accurate. For some parameters, no bound configuration was found in the explored regime of coupling constants.  
	The coupling constant variation for the $pp\Omega_{ccc}$ system is illustrated in Fig.~\ref{fig:ppOmega} and the corresponding three-body energies  are presented in Table~\ref{tab:ppOmega}, for without and with Coulomb interaction. 
	The corresponding complex-scaled spectrum is illustrated in Fig.~\ref{fig:ppOmega_csm}.
	
	As for the $nn\Omega_{ccc}$ system, the variation of the coupling constant yields no bound state in the $pp\Omega_{ccc}$ system, but possible resonances in the range $3-25$ MeV. This large range indicates lower accuracy of the extrapolation. However, the complex scaling does not reveal any resonances within this range. Therefore, similarly as above, also the $pp\Omega_{ccc}$ states most likely turn into virtual states at $\gamma=0$.
		
	\begin{table}
		\caption{Three-body energies (in MeV) of the $pp\Omega_{ccc}$ system. 
			the numbers between parenthesis correspond
			to the results with Coulomb interaction, Eq.~\eqref{eq:Coul}.  
		}
		\centering
		\begin{tabular}{c|cc c cc c cc }
			\hline
			Channel	& \multicolumn{2}{c}{$ 16 $}&& \multicolumn{2}{c}{$17$}&&  \multicolumn{2}{c}{$18$}\\
			\cline{1-1} \cline{2-3} \cline{5-6} \cline{8-9}
			$^{3}S_{1}$   & 2.95 & (11.59) && 4.40 &  (14.72) && 5.15 & (13.19) \\  
			$^{4}S_{3/2}$ & 4.22 & (11.84) && 4.59 &  (15.78) && 4.75 & (13.67) \\  
			$^{5}S_{2}$   & 4.10 & (12.49) && 5.81 &  (---)   && 5.36 & (---) \\  			
		\end{tabular}
		\label{tab:ppOmega}
	\end{table}	
	
	\subsection{$pn\Omega_{ccc}$ system}
\begin{figure*}[hbt!]
		\centering
		\includegraphics[scale=0.35]{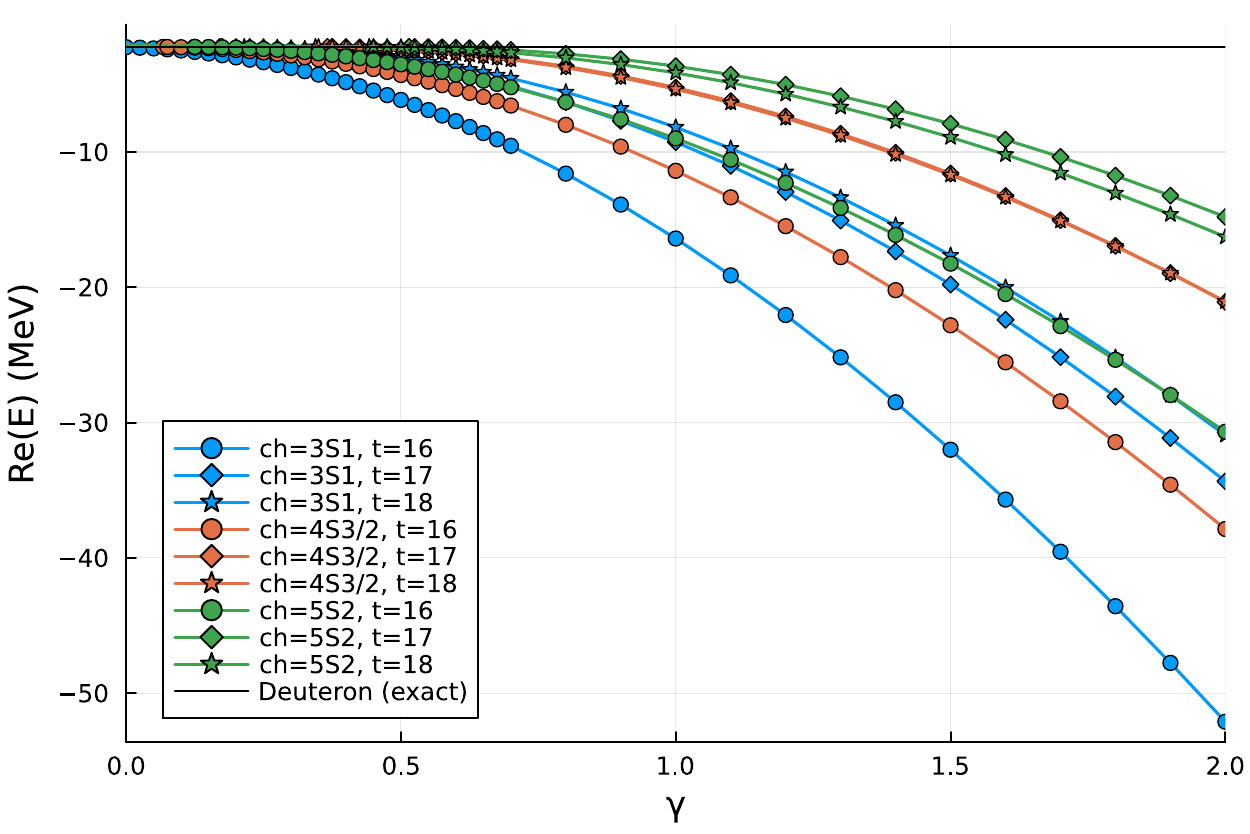} 	
		\includegraphics[scale=0.35]{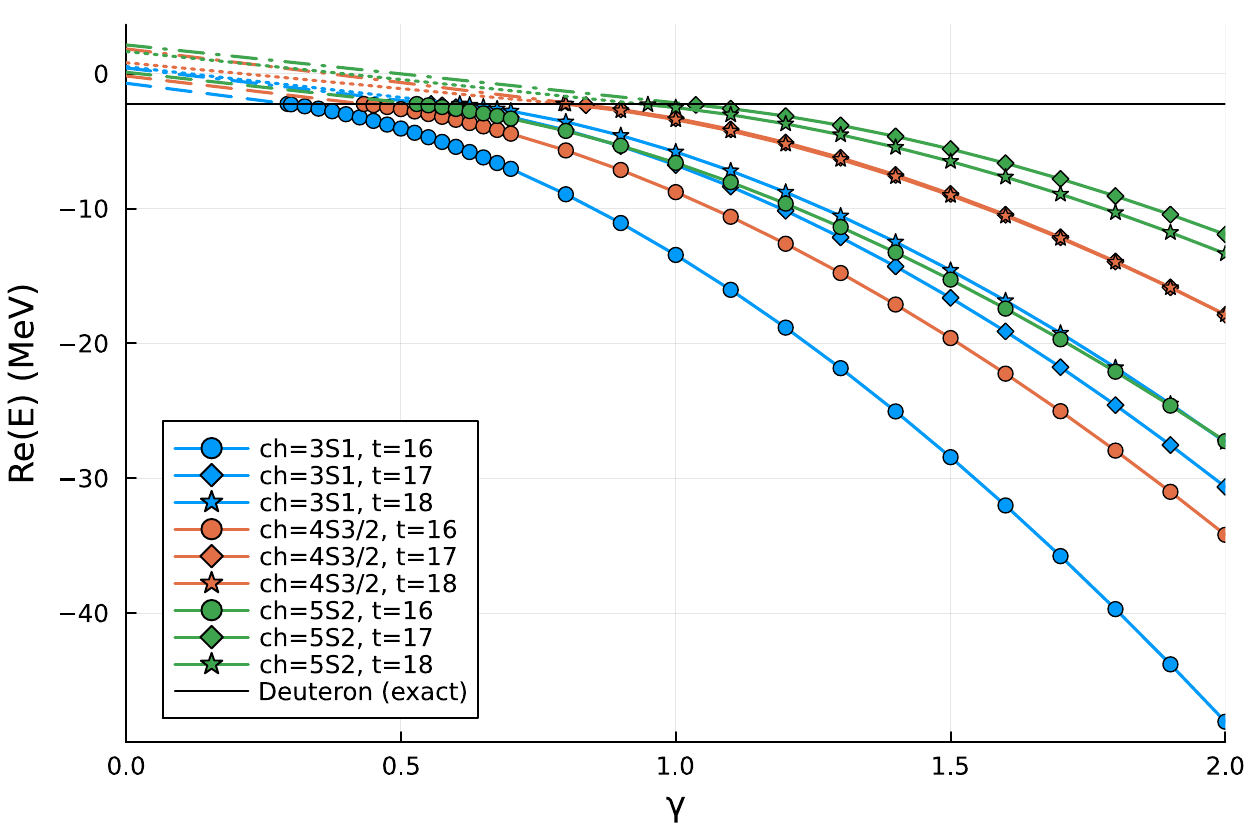}
		\caption{Coupling constant variation for the $pn\Omega_{ccc}$ system, without (left), and with (right) Coulomb interaction.\label{fig:pnOmega}}
	\end{figure*}

\begin{figure*}[hbt!]
		\centering
		\includegraphics[scale=0.4]{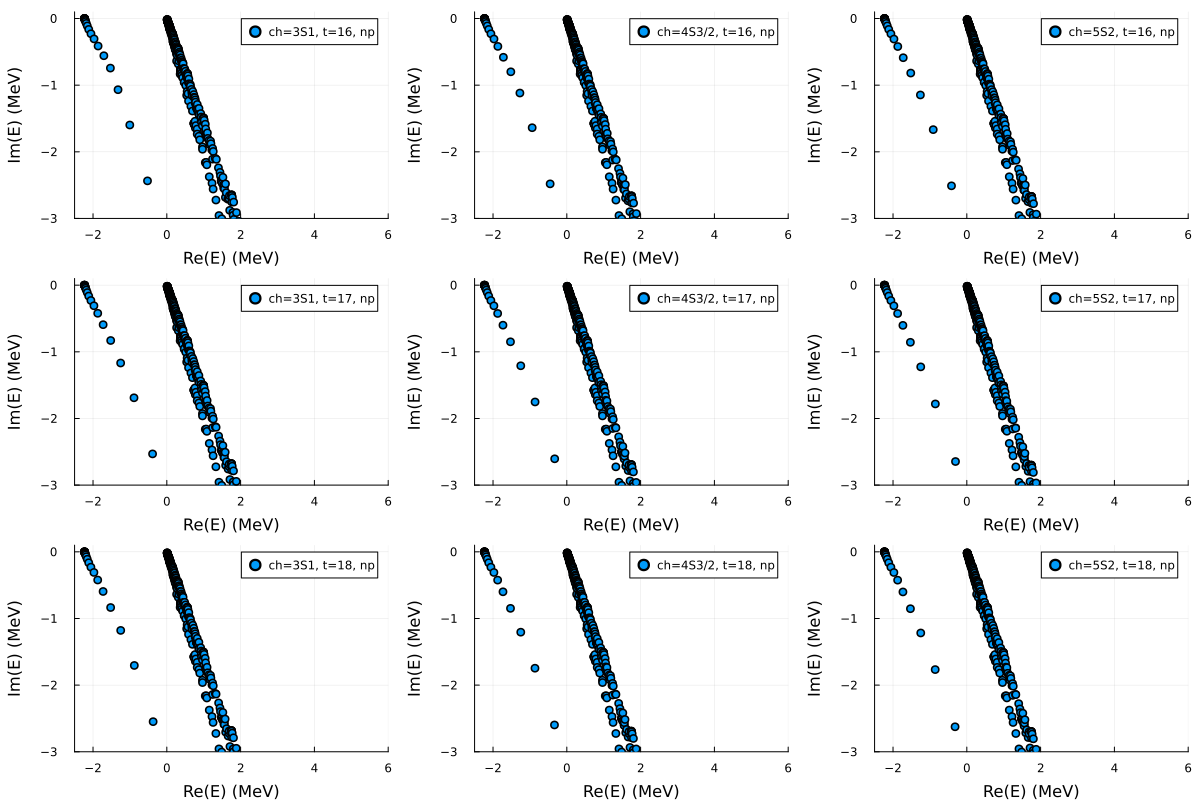}
		\caption{Complex-scaled spectrum for the $pn \Omega_{ccc}$  system. No resonance is found. 
			The $np$-subsystem is bound, hence we see two rotated continua, one starting at the deuteron binding energy, and one starting at the origin.			
			\label{fig:pnOmega_csm}}
	\end{figure*}

	 The lowest energies are observed in the $pn\Omega_{ccc}$ system, as anticipated due to the more attractive nature of the $np$ interaction featuring a bound state. 	
	The variation in the coupling constant of the $pn\Omega_{ccc}$ state is illustrated in Fig.~\ref{fig:pnOmega}, while the corresponding three-body energies are presented in Table~\ref{tab:pnOmega}.  
	Moreover, the corresponding complex-scaled spectrum is depicted in Fig.~\ref{fig:pnOmega_csm}. 
	If the figure is examined carefully, a very light “bump” near $-1$ can be seen in the rotated spectrum, and it is wondered whether some hidden structure, such as a very broad resonance, exists. However, we could not achieve convergence for larger rotation angles.

	\begin{table}
		\caption{Three-body energies (in MeV) of the $pn\Omega_{ccc}$ system. 	the numbers between parenthesis correspond
			to the results with Coulomb interaction, Eq.~\eqref{eq:Coul}. }
		\centering
		\begin{tabular}{c|cc c cc c cc }
			\hline
			Channel	& \multicolumn{2}{c}{$ 16 $}&& \multicolumn{2}{c}{$17$}&&  \multicolumn{2}{c}{$18$}\\
			\cline{1-1} \cline{2-3} \cline{5-6} \cline{8-9}
			$^{3}S_{1}$   & -2.25 & (-0.75) && -2.18 &  (0.40) && -2.14 &  (0.49) \\  
			$^{4}S_{3/2}$ & -2.20 & (-0.19) && -2.12 &  (1.83) && -2.15 &  (0.63) \\  
			$^{5}S_{2}$   & -2.21 & (-0.09) && -2.09 &  (2.12) && -2.12 &  (1.63) \\ 			
		\end{tabular}
		\label{tab:pnOmega}
	\end{table}	
	
	Across all analyzed $NN\Omega_{ccc}$ systems, we identify a three-body bound state only for the d-$\Omega_{ccc}$ state with spin $(I)J^{\pi} = (0)1/2^{+}$, $t/a=16$, and in the absence of the Coulomb interaction. The binding energy of this state, $B_{3} = -2.255$ MeV, lies below the deuteron binding energy ($B_d=-2.23$ MeV). The energy is well-converged and consistently remains below the deuteron binding energy.  It should be noted that the d-$\Omega_{ccc}$ state cannot couple to the lower channels $N \Lambda_{c}\Xi_{cc} $ and $N \Sigma_{c}\Xi_{cc} $, with the $ \Lambda_{c}\Xi_{cc} $ and the $ \Sigma_{c}\Xi_{cc} $  subsystems in S waves, hence the width of possible d-$\Omega_{ccc}$ near-threshold resonances is expected to be small.

	In Ref.~\cite{filikhin2025omega3cnn}, the existence of resonances in the d-$\Omega_{ccc}$ system was proposed, based on an approach involving the variation and continuation of the coupling constant. While qualitatively similar results were obtained in this study, significant quantitative discrepancies were observed. Our analysis demonstrates that for the d-$\Omega_{ccc}$ $(0)1/2^{+}$ case at $t/a=16$, the state is bound at $\gamma=0$ (the physical scenario). For other cases, the states appear to cross the deuteron binding energy, potentially forming near-threshold resonances, though these would be challenging to resolve.  
	These states may also become virtual states, however, in this case their computational identification remains unresolved.
	Possibly, differences in convergence criteria or interpolation methods in Ref.~\cite{filikhin2025omega3cnn} led to an overestimation of the energy at $\gamma=0$.

	\section{Conclusions and outlook\label{sec:Summary-and-conclusions}}
	Recent developments of interactions with physical quark masses in the charm sector based on lattice QCD, together with our results for the d-$\Omega_{ccc}$  system, are believed to advance the understanding of heavy-baryon interactions and may motivate experimental searches.
	
	Our complex scaling analysis suggests that the three-body states bound at large \(\gamma\) likely become virtual states rather than resonances at the physical value, analogous to certain Efimov states in atomic physics that cross the two-body threshold~\cite{naidon2017}. The complex scaling method provides indirect evidence for this interpretation which is not accessible through coupling constant variation alone.

	Furthermore, as discussed in Ref.~\cite{PhysRevX.14.031051}, the three-body interaction model results, such as those in~\cite{PhysRevC.108.064002}, are viewed as providing a more complete and apparently more precise description of ALICE data. Consequently, future femtoscopic studies in high-energy collisions may unveil the existence of these states. It is hoped that our theoretical studies will aid in the design of experiments where these lattice QCD–based predictions can be tested.				
	\section*{Acknowledgement}	
  L. H. is supported by the RIKEN special postdoctoral researcher program (SPDR).		
	
	\bibliography{Refs.bib}
	
\end{document}